\documentclass[aps,prl,showpacs,amsmath,amssymb,amsfonts,twocolumn,
superscriptaddress,floatfix,footinbib]{revtex4-1}

\usepackage{subfigure}
\usepackage{graphicx}
\usepackage[flushmargin]{footmisc}
\usepackage{color}
\usepackage{soul}
\usepackage{longtable}
\usepackage[normalem]{ulem}
\usepackage{amsmath}

\begin{document}

\title{Backscattering suppression in supersonic 1D polariton condensates}

\author{D. Tanese}
    \affiliation{CNRS, Laboratoire de Photonique et Nanostructures, Route de Nozay, 91460 Marcoussis, France}
\author{D. D. Solnyshkov}
   \affiliation{LASMEA, Clermont Universit\'e, University Blaise Pascal, CNRS, 24 avenue des Landais, 63177 Aubi\`ere cedex, France}
\author{A.~Amo}
\author{L.~Ferrier}
\author{E.~Bernet-Rollande}
\author{E.~Wertz}
\author{I.~Sagnes}
\author{A.~Lema\^itre}
\author{P.~Senellart}
   \affiliation{CNRS, Laboratoire de Photonique et Nanostructures, Route de Nozay, 91460 Marcoussis, France}
\author{G.~Malpuech}
   \affiliation{LASMEA, Clermont Universit\'e, University Blaise Pascal, CNRS, 24 avenue des Landais, 63177 Aubi\`ere cedex, France}
\author{J.~Bloch}
   \affiliation{CNRS, Laboratoire de Photonique et Nanostructures, Route de Nozay, 91460 Marcoussis, France}


\date{\today}

\begin{abstract}

We investigate
the effects of disorder on the propagation of one-dimensional
polariton condensates in semiconductor microcavities. We observe a strong suppression of the backscattering produced by the imperfections of
the structure when increasing the condensate density. This 
suppression occurs in the supersonic regime and is simultaneous to the onset of parametric instabilities
which enable the "hopping" of the condensate through the disorder.
Our results evidence a new mechanism for the frictionless flow of
polaritons at high speeds.

\end{abstract}

\maketitle

The interplay between kinetic energy, localization energy, and particle interactions lies
at the heart of the transport properties in condensed matter physics. This problem
has been thoroughly treated theoretically since the seminal work of Anderson~\cite{Anderson1958, Lagendijk2009},
 which described the localization of electrons in disordered media caused by the interference between the electronic wavefunctions.

  \begin{figure*}[t]
          \centering
  \includegraphics[width=1\textwidth]{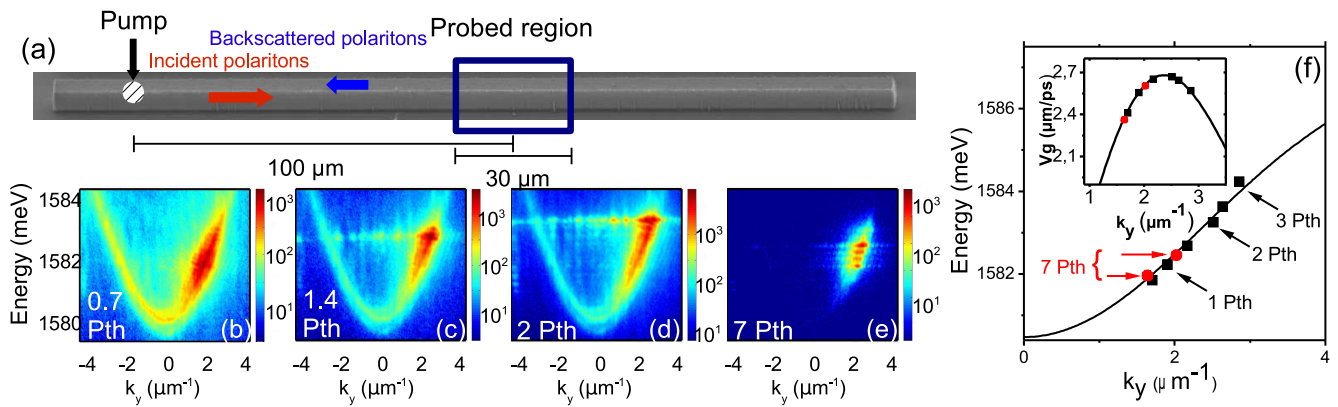}

  \caption{(Color online) (a) Scanning electron microscope image of the wire and scheme of the experiment. (b)-(e) Normalised far field emission of the polariton gas in the probed region. (f) Energy and momentum of the polariton condensates at different excitation powers, below (black squared points) and above (red round points) the onset for stimulated parametric processes. The solid line shows the polariton dispersion and the inset the corresponding values of the group velocity.}\label{Fig1}
  \end{figure*}

Electron localization phenomena have been observed in doped semiconductors~\cite{Rosenbaum1983}, but this systems suffers from  a large number of non-controllable parameters, like the strong
Coulomb interactions or the disorder potential. As an alternative,
gases of neutral bosons appear as model systems for the understanding of localization. While
some works have addressed these questions in the eighties~\cite{Fischer1989}, this activity
took an enormous theoretical and experimental expansion since the observation of Bose-Einstein
 condensates of ultracold atoms~\cite{Anderson1995,*Davis1995}, and has allowed the study of localization at low dimensions
 in controlled disorder landscapes. One dimensional systems are of particular interest as the
  reduction of the available scattering channels enhances the interference effects. Formally,
  this gives rise to localization for any disorder strength~\cite{Mott1961b}. Recent experiments
   have indeed shown the localization of bosonic matter waves in the presence of random disorder using ultracold
   atoms~\cite{Billy2008, Roati2008}.

While disorder tends to localize the wavepackets, interparticle repulsive interactions have a delocalization
 effect. A localized to extended phase transition has been predicted to be triggered by the repulsion
  between particles in 1D~\cite{Paul2005,*Paul2009, Aleiner2010}, and recent experiments have
   demonstrated this effect in	 atomic condensates at rest~\cite{Deissler2010}.
Particularly interesting is the case of a 1D boson flow in the presence of a weak disordered potential. If the considered sample is small enough
   and/or the disorder is weak, a quasi-extended state is recovered
   with a high transmission~\cite{Pendry1994},
    still with an appreciable backscattering amplitude. 
Repulsive interactions in such a flowing boson gas
    could lead to the formation of a superfluid state, resulting in the perfect transmission through the 1D
    channel~\cite{Paul2005,*Paul2009}. 
However, the superfluid behaviour is lost if the flow speed becomes larger than a critical velocity~\cite{Ianeselli2006} which
is close to the speed of sound of the fluid given by $c=\sqrt{\alpha n/m}$, where $m$ is the mass of the particles, $n$ is their density, and $\alpha$ the interparticle interaction energy.
Nevertheless, as we will see in this work, even at high flow speeds in the non-superfluid regime, interactions between particles can lead to a strong reduction of the backscattering 
and, consequently, to an enhancement of the transmission.

In this letter we study the effects of interparticle interactions
on the propagation of a polariton boson condensate in a 1D semiconductor wire microcavity. We observe, in 
the supersonic regime, a strong suppression of the signal backscattered by the imperfections of the structure 
while increasing the condensate density. This suppression, 
which was previously theoretically predicted in ~\cite{Paul2005,*Paul2009},
is caused by the onset of parametric instabilities arising from particle interactions, giving the condensate access to propagating states which bypass the potential barriers caused by disorder. This confirms experimentally the existence of a new mechanism for the suppression of the backscattering, very different from the recently reported superfluidity of polaritons~\cite{Carusotto2004, Amo2008}.

Cavity polaritons are bosonic particles arising from the
strong coupling between quantum well excitons and photons
 confined in a microcavity. Their Bose-Einstein condensation has 
 been reported by several groups~\cite{Kasprzak2006, Balili2007, 
 Christopoulos2007,Wertz2009}. 
 Non-linear phenomena such as optical parametric oscillation 
(OPO)~\cite{Stevenson2000,Diederichs2006},
multistable behaviour~\cite{Paraiso2011}, and superfluidity ~\cite{Amo2009, Amo2008} have been demonstrated.
Recent technological breakthroughs have allowed the production of
very high quality microcavities 
with polariton lifetimes as long as 30~ps -5~times larger than in the pre-existing 
samples. In a 1D geometry, this achievement has allowed the 
observation of propagating polariton condensates spatially 
separated from the pumping area and showing phase coherence 
extending over one hundred healing lengths~\cite{Wertz2010}. Here we use this geometry to study the effect of interactions on the backscattering caused by imperfections in the structure.

Our samples are GaAs/AlGaAs based microcavities etched into 1D wires
with a width of 3.5~$\mu$m and a length of 0.2~mm~\cite{Wertz2010}.
The Rabi splitting amounts to 15~meV and the
polariton lifetime to 30~ps. Polariton condensates
are formed under non-resonant photoexcitation (laser
wavelength around 730~nm) with a single-mode Ti:Sapph laser.
The laser spot has a 2~$\mu$m
diameter and is positioned close to one of the edges of the wire [see Fig.~\ref{Fig1}(a)].
At low excitation density, carriers are photoinjected at the position of the excitation spot,
relaxing down to form polaritons which populate the lower polariton branch.
 This incoherent gas of polaritons with different momenta can propagate long distances
 in the wire thanks to the long cavity lifetime. Figure~\ref{Fig1}(b) shows the momentum
 space in logarithmic scale of the polariton signal 100~$\mu$m away from the injection area.
For positive values of the momentum, the forward propagating polaritons can be observed. Due to the large value of the
exciton mass, the population of the excitonic reservoir is 
negligible out of the excitation area.

\begin{figure}[t!]
  \includegraphics[width=1\columnwidth]{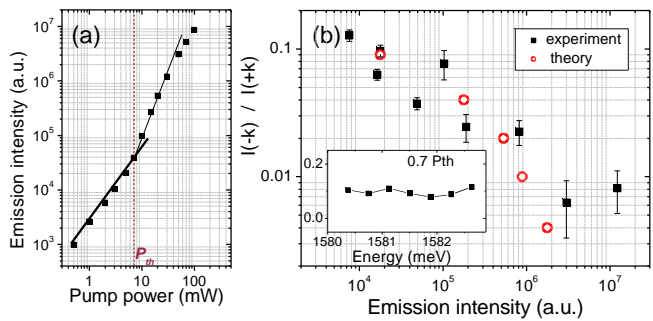}
  \caption{(Color online) (a) Emitted intensity in the probed region as a function of excitation power. (b) Black points (red circles): ratio of the backscattered [$I(-k)$] to the incident polariton signal [$I(+k)$] in the probed region measured at the energy of the peak signal, experimentally (theoretically) obtained in the conditions of Fig.~\ref{Fig1}. Inset: backscattered ratio measured at low excitation power (below the condensation threshold) as a function of the energy of the propagating polaritons.}\label{Fig2}
  \end{figure}

Above a threshold excitation density
$P_{th}$ [Fig.~\ref{Fig2}(a)], polariton condensation takes place at $k=0$ in the region
optically pumped by the laser. In that region, the presence of a
dense excitonic reservoir induces a strong blueshift of the
polariton energy even at few times $P_{th}$, as discussed in
Ref.~\cite{Wertz2010}. The energy difference between the $k=0$
states \emph{in} and \emph{out} of the pumped area results in an 
acceleration of the polariton condensate from the pump spot. As a 
result, in the region of observation, the propagating 
condensate preserves the same emission energy as in the pumped 
region but it gains a finite momentum [2.1~$\mu$m$^{-1}$ in
Fig.~\ref{Fig1}(c)]. By further increasing the polariton density,
the reservoir in the pumped area induces a larger blueshift and
the propagating condensate emits at higher energy with a larger $k$,
as shown in (d) and in the black dots of Fig.~\ref{Fig1}(f).

This progression continues while increasing the excitation density
until the propagating condensate gets close to the inflexion point
of the lower polariton branch.
Then, parametric instabilities induced by polariton interactions set up
the spontaneous oscillation of the polariton emission between
several coexisting states with different
energies [see Fig.~\ref{Fig1}(e)]. Note that contrary to the case in non-linear crystals or in previous OPO experiments in 
polaritons~\cite{Stevenson2000,Diederichs2006}, none of the states in the OPO is resonantly fed by a laser source. Here, the OPO is spontaneously triggered once a sufficiently high population is accumulated in the propagating condensate. 
This is a remarkable example of a self-sustained optical parametric oscillator. As we will see below, these parametric
instabilities play an important role in the quenching of the polariton backscattering.

Let us now concentrate on the scattering of the propagating polariton gas with the weak disorder present
 in the wire microcavity. The origin of the disorder is structural and gives rise to a change of the polariton
  energy smaller than 0.1-0.4~meV. It is mainly caused by the weak
 strain accumulated during the growth of the mirrors and also to 
 the non-intentional rugosity that is formed on the walls of the 
 wires during the lithographic and etching processes. 
 Figure~\ref{Fig1}(b) shows a significant signal backscattered
 from the incoherent polariton gas at negative momenta. 
The origin of this signal in the far field can only
   arise from the scattering of propagating polaritons with disorder as those states cannot be
fed by the excitonic reservoir, located 100~$\mu$m away from the 
observation area. When the polariton density is increased, the 
relative backscattered signal is strongly reduced, as 
shown in Fig.~\ref{Fig1}(b)-(e) and summarized in 
Fig.~\ref{Fig2}(b). The quenching of the backscattering reaches 
a factor of 20 at the highest investigated excitation density, with 
respect to the low density regime.

This remarkable reduction is not caused by the increase of the kinetic energy with excitation power, since the intrinsic backscattered amplitude, measured at low density, does not depend on the polariton energy [see inset of Fig.~\ref{Fig2}(b)]. The origin of the strong reduction of the scattering can neither be in a transition to a superfluid regime at  high density in the context of the Landau criterion, as recently reported in two dimensions under resonant excitation~\cite{Carusotto2004,Amo2008}. Our condensates travel at large speeds, with
 momenta on the order of 2-3~$\mu$m$^{-1}$ and energy $\sim$3~meV above the $k=0$ state. In order
to reach the superfluid regime, interaction induced blueshifts at 
least twice the value of the kinetic energy would be required. We are far from that situation, meaning that our fluids
are supersonic.

\begin{figure}[t!]
  \includegraphics[width= 1\columnwidth]{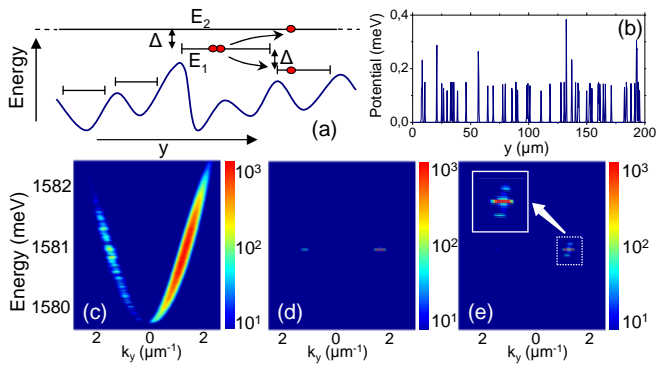}
  \caption{(Color online) (a) Schematic representation of the disorder potential showing the spatial extension of localised states (horizontal bars) and the parametrically induced hopping process. (b) Disorder potential used in the model. (c)-(e) Simulation of the polariton far field emission in the disordered potential with increasing polariton density. The upper left squared region in (e) shows a magnification of the lower squared area.}
  \label{Fig3}
  \end{figure}

We propose a novel mechanism for the quenching of the backscattering based on the onset of spontaneous
parametric processes as the polariton density is increased. In order to understand this mechanism,
let us go back to the Anderson model of localization by disorder~\cite{Anderson1958}.
One of its simplest descriptions is to consider a regular 1D lattice in the tight-binding model with a
 hopping constant $J$, the energy of each site being random [see Fig.~\ref{Fig3}(a)]. If the energy
 difference between two sites ($E_1, E_2$) is smaller than $J$ ($\left|E_1-E_2\right|<J$), the particle
 can jump from one site to the other. If the probability of finding an available neighbor for the jump
 is $P$ ($<1$), then the probability to make $N$ jumps will be $P^N$, which gives an
 exponentially decaying distribution around the initial site.

Let us consider now a system with local repulsive interaction between particles. This interaction can shift the on-site energy, increasing the probability to match the energies of two different sites. 
Additionally, the onset of OPO on-site scattering can populate virtually any energy state via the following energy conserving process: $E_1+E_1\rightarrow(E_1+\Delta)+(E_1-\Delta)$, where $2\Delta$ is the energy difference between signal and idler states. If the condition $\left|E_1\pm\Delta-E_2\right|<J$ is verified, a particle can jump to the neighboring site [Fig.~\ref{Fig3}(a)]. 

In our experiment the kinetic energy of particles is much larger than the localization energy. Therefore, the effect of local blueshift is expected to be 
weak compared to that of the onset of parametric instabilities, which increases the probability of finding a neighbor with a matching energy and tends to suppress localization.
This phenomenon has been theoretically discussed in Ref.~\cite{Aleiner2010} in the context of ultracold atoms, while a similar effect has been addressed within a different approach in Ref.~\cite{Paul2005,*Paul2009}. In the latter case, calculations showed that because of the interactions the decay of the transmission of a 1D atomic Bose-Einstein condensates through a disordered waveguide is algebraic instead of exponential, a signature of the suppression of Anderson localization. This regime corresponds to "non-stationary" solutions of the Gross-Pitaevskii equation, evidencing the presence of many-energy states in the system, in line with our interpretation of the onset of parametric processes in experiments.

In order to support our model for the quenching of the
backscattering, we have performed numerical simulations of the
propagation of polariton condensates with a certain wave-vector
through a series of point-like defects. To describe accurately the
polariton condensate, we use a set of Schr\"{o}dinger
equations describing the time evolution of the photonic
$\psi_{ph}(x,t)$ and
 excitonic $\psi_{ex}(x,t)$ mean fields coupled via the light matter interaction (Rabi splitting, $\Omega_R=15$ meV).
 Our model takes into account the two allowed spin projections of 
the photon and exciton fields $\sigma=\pm1$ \cite{Shelykh2006}:

\begin{eqnarray}
i\hbar \frac{{\partial \psi _{ph}^\sigma }}{{\partial t}} &=&  - \frac{{{\hbar ^2}}}{{2{m_{ph}}}}\Delta \psi _{ph}^\sigma + \frac{\Omega_R }{2}\psi _{ex}^\sigma \\
\nonumber &-& \frac{{i\hbar }}{{2{\tau _{ph}}}}\psi _{ph}^\sigma  + {P^\sigma } + H_{eff}\psi _{ph}^{ - \sigma }+U\psi _{ph}^\sigma\\
i\hbar \frac{{\partial \psi _{ex}^\sigma }}{{\partial t}} &=&  - \frac{{{\hbar ^2}}}{{2{m_{ex}}}}\Delta \psi _{ex}^\sigma + \frac{\Omega_R }{2}\psi _{ph}^\sigma \\
\nonumber &-& \frac{{i\hbar }}{{2{\tau _{ex}}}}{\psi _{ex}^\sigma} +  \alpha\left| \psi _{ex}^\sigma \right|^2\psi _{ex}^\sigma
\end{eqnarray}

Here $m_{ph}=3.6\times10^{-5} m_0$, $m_{ex}=0.4 m_0$ and $m_0$ are
the cavity photon, the quantum well exciton and the free electron
masses, respectively. The lifetime of the particles are
$\tau_{ph}=15$~ps and $\tau_{ex}=400$~ps. 
$H_{eff}$ accounts for the TE-TM polarization splitting
of the waveguided modes.
$\alpha=6E_{b}a_{B}^2/S$ is the polariton-polariton interaction
constant ($E_{b}=6$~meV is the exciton binding energy, $a_B=10$~nm
is the exciton Bohr radius), $U(x)$ is the disorder potential along
the wire which is a series of randomly spaced delta peaks [shown in Fig.~\ref{Fig3}(b)], $P(x,t)$ is the coherent $cw$ pumping, localized on a spot
of 2~$\mu$m placed on the left of the sample, as sketched on
Fig.~\ref{Fig1}(a), injecting polaritons with well defined
momenta. 

Figures~\ref{Fig3}(c)-(e) show the far field emission 100~$\mu$m
away from the excitation spot obtained from the simulations at
different excitation densities. To simulate the situation below
threshold [Fig.~\ref{Fig3}(c)], we inject a short
pulse, filling all positive momenta states of the lower polariton
branch with a distribution similar to the measured one [Fig.~\ref{Fig1}(b)]. A strong backscattered signal can be
observed, analogous to the experimental one shown in
Fig.~\ref{Fig1}(b). Note that the backscattering is entirely due to
the presence of the disorder potential. Just above threshold the
condensate presents a single energy [Fig.~\ref{Fig1}(c)], which we simulate by quasiresonant $cw$ pumping with a well defined $k$ [Fig.~\ref{Fig3}(d)]. A significant
backscattered signal is still present. At higher densities
[Fig.~\ref{Fig3}(e)], new propagation frequencies associated with
the onset of parametric processes become clearly visible, whereas
the pumping is still monochromatic. The well-defined separation
between the frequencies, which can be observed both in the experimental [Fig.~\ref{Fig1}(d)] and theoretical images [Fig.~\ref{Fig3}(e)], is due to the polarization splitting $2H_{eff}\approx 0.5$~meV. It determines the preferential resonances, for the onset of the OPO processes.
This model provides an interpretation to the observation of
apparently spectrally equidistant condensates~\cite{Wertz2010} at
high pumping rates. The energy spacing between the different
frequencies is in our approach not related to a balance between
scattering rates and losses~\cite{Wouters2010b}, but to the value of the TE-TM splitting.

Similar to the results of Ref.~\cite{Paul2005,*Paul2009}, the appearance of the new frequencies results in the reduction
of the backscattered signal.
Figure~\ref{Fig2}(b) summarizes
 in open dots the calculated backscattered intensity averaged over 50 disorder realisations, as a function of the excitation density. 
 They show a  reduction by more than one order of magnitude when the density is increased by a factor of 100, in good
 agreement with the experimental results. 
This value is significantly greater than the one found in experiments of superfluidity 
in planar microcavities (a quenching by a factor of 4)~\cite{Amo2008}. Though backscattering effects are less important in 2D than in 1D systems due to the very different density of available states in each case, the mechanism we propose
might as well play a role on the flow of polaritons in triggered optical parametric configurations in planar microcavities~\cite{Amo2009, Sanvitto2009}.

In conclusion, we have observed a strong suppression of the
scattering
 in propagating
 1D polariton condensates caused by the onset of parametric instabilities at high excitation densities,
  which enable the "hopping" of the condensate through the disorder. This new mechanism results in a
  quasi-frictionless flow of supersonic polariton condensates, opening the way to the fabrication of
  integrated polariton circuits with high transmissivity.

We are greatful to A.~V.~Kavokin, P.~Leboeuf and N.~Pavloff for fruitful discussions. This work was partly supported by the C'Nano Ile de France contract "Sophiie2", the RTRA contract "Picorre", and the FP7 ITNs "Clermont4" (235114) and "Spin-Optronics"(237252).



%

\end{document}